\documentclass[%
 reprint,
superscriptaddress,
amsmath,amssymb,
aps,
pra,
]{revtex4-1}
\usepackage{caption}
\usepackage{subcaption}
\usepackage{graphicx}
\usepackage{xcolor}
\usepackage{graphicx}
\usepackage{dcolumn}
\usepackage{bm}
\usepackage{upgreek}
\usepackage{lineno}

\begin{document}

\preprint{APS/123-QED}
\title{Ultra-compact synthesis of space-time wave packets}

\author{Murat Yessenov}
\thanks{Corresponding author: yessenov@knights.ucf.edu}
\affiliation{CREOL, The College of Optics \& Photonics, University of Central Florida, Orlando, FL 32816, USA}
\author{Oussama Mhibik}
\affiliation{CREOL, The College of Optics \& Photonics, University of Central Florida, Orlando, FL 32816, USA}
\author{Lam Mach}
\affiliation{CREOL, The College of Optics \& Photonics, University of Central Florida, Orlando, FL 32816, USA}

\author{Tina M. Hayward}
\affiliation{Department of Electrical and Computer Engineering, University of Utah, Salt Lake City, Utah 84112, USA}

\author{Rajesh Menon}
\affiliation{Department of Electrical and Computer Engineering, University of Utah, Salt Lake City, Utah 84112, USA}

\author{Leonid Glebov}
\affiliation{CREOL, The College of Optics \& Photonics, University of Central Florida, Orlando, FL 32816, USA}
\author{Ivan Divliansky }
\affiliation{CREOL, The College of Optics \& Photonics, University of Central Florida, Orlando, FL 32816, USA}
\author{Ayman F. Abouraddy}
\thanks{raddy@creol.ucf.edu}
\affiliation{CREOL, The College of Optics \& Photonics, University of Central Florida, Orlando, FL 32816, USA}

\begin{abstract}
Space-time wave packets (STWPs) are pulsed fields in which a strictly prescribed association between the spatial and temporal frequencies yields surprising and useful behavior. However, STWPs to date have been synthesized using bulky free-space optical systems that require precise alignment. We describe a compact system that makes use of a novel optical component: a chirped volume Bragg grating that is rotated by $45^{\circ}$ with respect to the plane-parallel device facets. By virtue of this grating's unique structure, cascaded gratings resolve and recombine the spectrum without free-space propagation or collimation. We produce STWPs by placing a phase plate that spatially modulates the resolved spectrum between such cascaded gratings, with a device volume  of $25\!\times\!25\!\times\!8$~mm$^{3}$, which is orders-of-magnitude smaller than previous arrangements.
\end{abstract}



\maketitle

Unlike conventional pulsed beams whose spectra are separable with respect to the spatial and temporal degrees of freedom, space-time wave packets (STWPs) \cite{Yessenov22AOP} are structured pulsed fields in which a tight association is enforced between the spatial and temporal frequencies \cite{Turunen10PO,FigueroaBook14}. Unique features of STWPs stem from this prescribed spatio-temporal spectral structure, including propagation invariance in linear media \cite{Malaguti08OL,Kondakci17NP,Porras17OL}, self-healing \cite{Kondakci18OL}, tunable group velocity \cite{Salo01JOA,Wong17ACSP2,Efremidis17OL,Kondakci19NC}, and anomalous refraction \cite{Bhaduri20NP}. Unlike previously theorized propagation-invariant focus-wave modes \cite{Brittingham83JAP} and X-waves \cite{Saari97PRL}, which did not make experimental headway \cite{Turunen10PO}, STWPs are readily synthesizable without need for exorbitantly broad spectra nor large numerical apertures \cite{Yessenov22AOP}. Moreover, STWPs are part of a burgeoning effort investigating spatio-temporally structured fields, including flying-focus pulses \cite{SaintMarie17Optica,Froula18NP,Jolly20OE}, wave packets with transverse angular momentum \cite{Hancock19Optica,Chong20NP,Gui21NP}, toroidal pulses \cite{Wan22NP,Zdagkas22NP}, among other examples \cite{Chen22SA,Yessenov22AOP}. A shared feature in the synthesis of such fields is that the initial pulse spectrum is spatially resolved via a diffraction grating to spatially modulate each wavelength. Consequently, these optical arrangements are bulky and require careful alignment.

Chirped Bragg volume gratings (CBGs) offer an alternative route to spatially resolving the spectrum \cite{Glebov14OEng}. At normal incidence, a CBG stretches the reflected pulse temporally \cite{Kaim14OEng}, but at oblique incidence the reflected wavelengths are spatially resolved, which can be used in multiplexing and demultiplexing in optical communications systems \cite{Gerken03IEEEPTL}. Recently, we made use of such a component to produce linear spatial chirp without the spectral phase introduced by conventional gratings \cite{Yessenov22NC,Yessenov22OL}. Nevertheless, CBGs typically have a large footprint because the reflected field exits from the same input facet, so that increasing the bandwidth requires increasing both the length \textit{and} the width of the device (e.g., CBG footprint in \cite{Yessenov22NC,Yessenov22OL} was 37$\times$25$\times$6 mm$^{3}$).

We recently introduced a new optical component that lifts these conventional restrictions: a \textit{rotated} chirped Bragg volume grating (r-CBG), in which the volume grating structure is rotated by $45^{\circ}$ with respect to the plane-parallel device facets \cite{Mhibik23OL}. Such a device has several salutary features: (1) the input beam is incident normally on one facet, and the spectrally resolved field exits a different facet also normally; (2) the emerging spectrum is collimated without requiring a lens and does not require further free-space propagation; (3) the link between the length and width of the device is severed, so that its cross-section can take the form of a narrow rectangle rather than a large square; and (4) a pair of r-CBGs can be cascaded to resolve and then recombine the spectrum, thus reconstituting the input pulse at the output.

Here, we construct an ultra-compact optical system for synthesizing STWPs comprising a cascaded pair of r-CBGs. The configuration is based conceptually on that in our previous work employing transmissive phase plates for spatially modulating the spectrally resolved wavefront of an ultrafast pulse produced by diffraction gratings \cite{Kondakci18OE}. Here, we sandwich such a phase plate between two appropriately cascaded r-CBGs to spatially modulate the spectrum with no need for any free-space propagation or collimation. This new spatio-temporal synthesis system has an ultra-compact footprint ($25\times25\times8$~mm$^{3}$) whose volume is orders-of-magnitude smaller with respect to that in \cite{Kondakci18OE}, is easily aligned, and is significantly more robust, thus offering opportunities for producing STWPs in moving platforms.

\begin{figure}[t!]
    \centering
    \includegraphics[width=8.6cm]{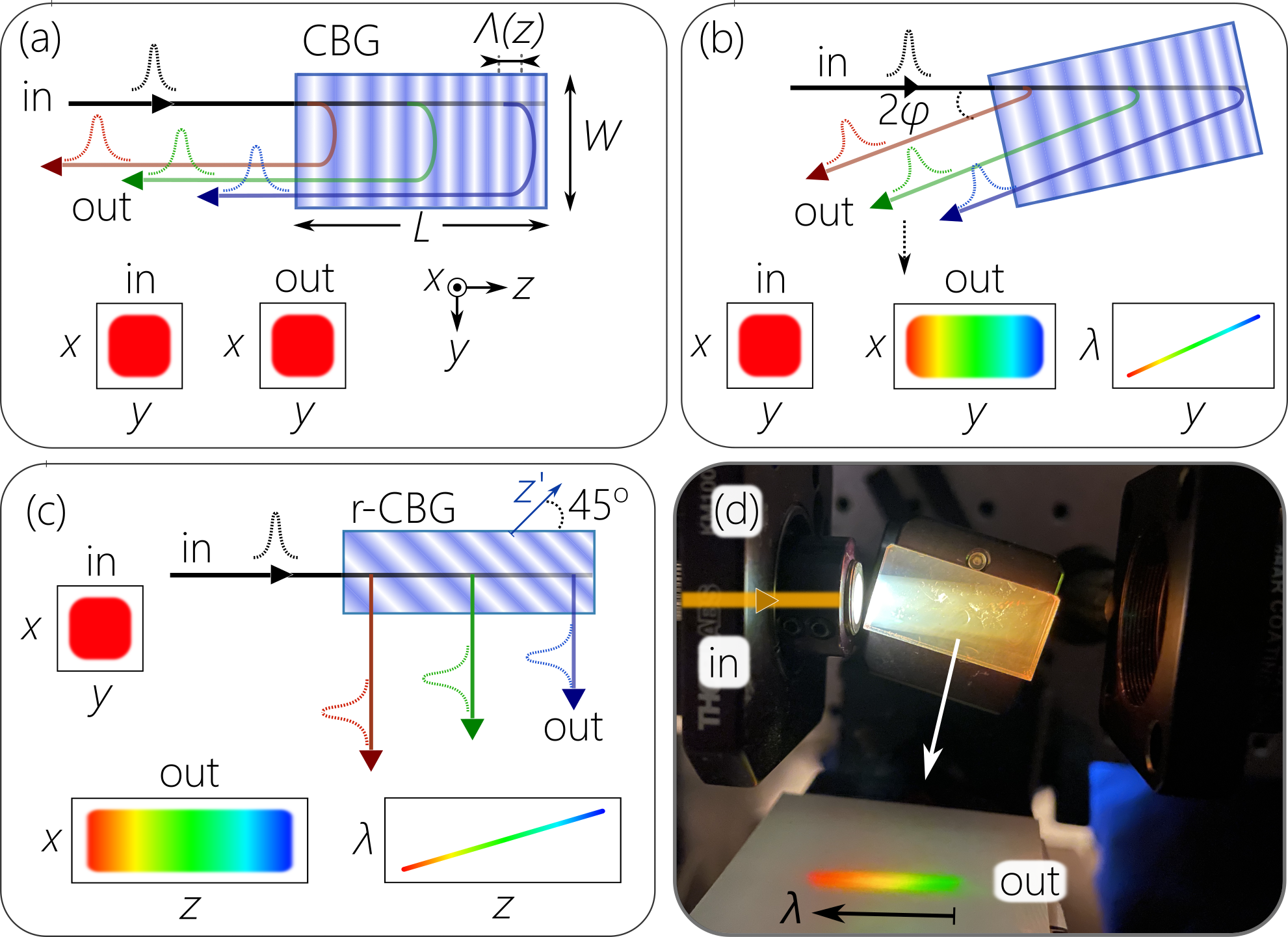}
    \caption{Concept of an r-CBG. (a) A pulse normally incident on a CBG is temporally stretched but not spectrally resolved. (b) A pulse obliquely incident on a CBG is spectrally resolved. (c) Spectrally resolving a pulse normally incident on an r-CBG. (d) Photograph of an r-CBG resolving the spectrum of white light. The r-CBG is designed for operation at $\lambda_{\mathrm{o}}\!\approx\!800$~nm, so operation in the visible requires oblique incidence.}
    \label{fig:RotatedVBGconcept}
\end{figure}

The basic concept of an r-CBG is illustrated in Fig.~\ref{fig:RotatedVBGconcept}. Traditionally, a pulse normally incident on a conventional CBG \cite{Glebov14OEng,Kaim14OEng} (in which the Bragg structure is parallel to the input facet) is temporally stretched [Fig.~\ref{fig:RotatedVBGconcept}(a)]. At oblique incidence, each wavelength reflects from a different depth within the CBG, and spatial dispersion resolves the spectrum of a collimated pulse, which is reflected back obliquely from the same input facet [Fig.~\ref{fig:RotatedVBGconcept}(b)]. It can be shown that the ratio of the CBG width $W$ and length $L$ is $\tfrac{W}{L}\!=\!2\tfrac{\sin\varphi}{\sqrt{n_{\mathrm{o}}^{2}-\sin^{2}\varphi}}$, where $\varphi$ is the external incident angle with respect to the normal to the CBG entrance facet, $n_{\mathrm{o}}$ is the average refractive index. Increasing the bandwidth at a fixed chirp rate requires increasing $L$, which therefore implies simultaneously increasing its width $W$ \cite{Mhibik23OL}. 

The change in geometry associated with rotating the Bragg structure by $45^{\circ}$ with respect to the plane-parallel device facets (r-CBG) produces a new configuration for spectral analysis. Here, the collimated pulse is incident \textit{normally} on the input facet (width $W$), and the resolved spectrum exits from the orthogonal facet (length $L$) also normally [Fig.~\ref{fig:RotatedVBGconcept}(c)]. This convenient configuration simplifies the alignment and the cascading of further optical elements (including other r-CBGs). Moreover, the width $W$ and length $L$ are now independent of each other: $L$ can be extended to increase the resolved bandwidth while holding $W$ fixed. The dimensions $L\times W$ for an r-CBG can thus be smaller than those for a CBG of the same chirp rate and bandwidth. In our experiments, we make use of r-CBGs of dimensions $25\times12.5\times8$~mm$^{3}$ designed for operation in the vicinity of $\lambda_{\mathrm{o}}\!\sim\!800$~nm. The dimension 12.5~mm can be reduced to $\sim\!4$~mm without a change in the device performance. A photograph of the r-CBG is shown in Fig.~\ref{fig:RotatedVBGconcept}(d), where it is obliquely oriented to operate in the visible.

\begin{figure}[t!]
    \centering
    \includegraphics[width=8.6cm]{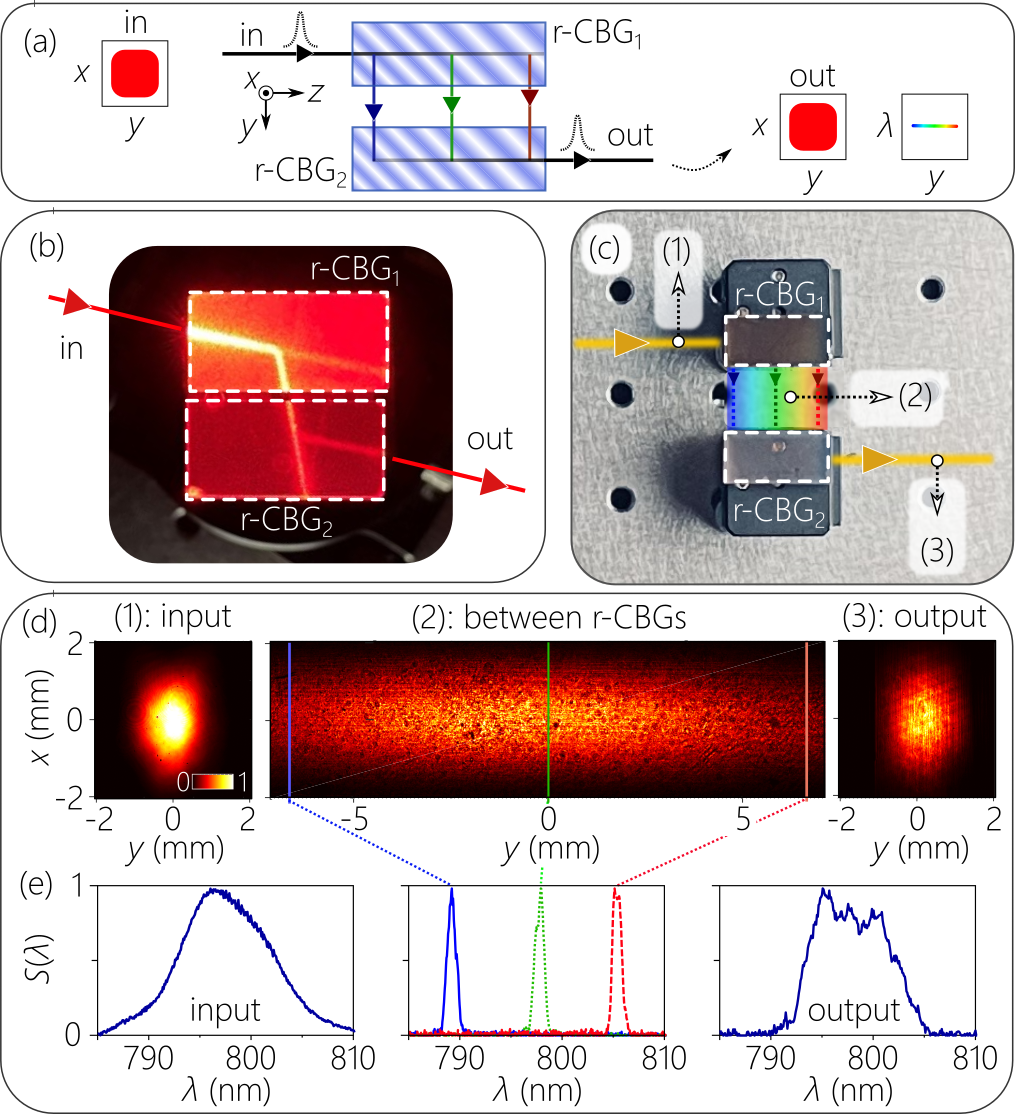}
    \caption{(a) Cascaded r-CBGs resolve and then recombine the spectrum. (b) Photograph of cascaded r-CBGs with a monochromatic red input laser. (c) Schematic of the cascaded configuration, where we capture (d) the intensity profile and (e) the spectrum at 3 different locations: (1) at the input, (2) after r-CBG${_1}$, and (3) after r-CBG${_{2}}$.}
    \label{fig:DoubleCBG}
\end{figure}

A pair of r-CBGs [Fig.~\ref{fig:DoubleCBG}(a)] can be cascaded to resolve and then recombine the spectrum. A collimated pulse is incident normally on the first (r-CBG$_{1}$) on the facet of width $W$, and the spatially resolved spectrum emerges from the facet of length $L$. This spectrum then impinges on the corresponding facet of the appropriately oriented second device (r-CBG$_{2}$), thereby combining the spectrum and reconstituting the pulse that emerges from the facet of width $W$. Physical realizations are shown in Fig.~\ref{fig:DoubleCBG}(b,c). We confirm that the input beam impinging on r-CBG$_{1}$ and the output beam emerging from r-CBG$_{2}$ have the same spatial structure [Fig.~\ref{fig:DoubleCBG}(d)] and the same spectrum [Fig.~\ref{fig:DoubleCBG}(e)]. Between r-CBG$_{1}$ and r-CBG$_{2}$, the spectrum has been spread over a large spatial width [Fig.~\ref{fig:DoubleCBG}(d,e)]. In (d-e) the input source is a mode-locked Ti:Sapp laser having 10-nm-bandwidth (FWHM), centered at a wavelength $\approx$800~nm, and a beam width of $\sim\!1.5$~mm. 

\begin{figure*}[t!]
    \centering
    \includegraphics[width=17.8cm]{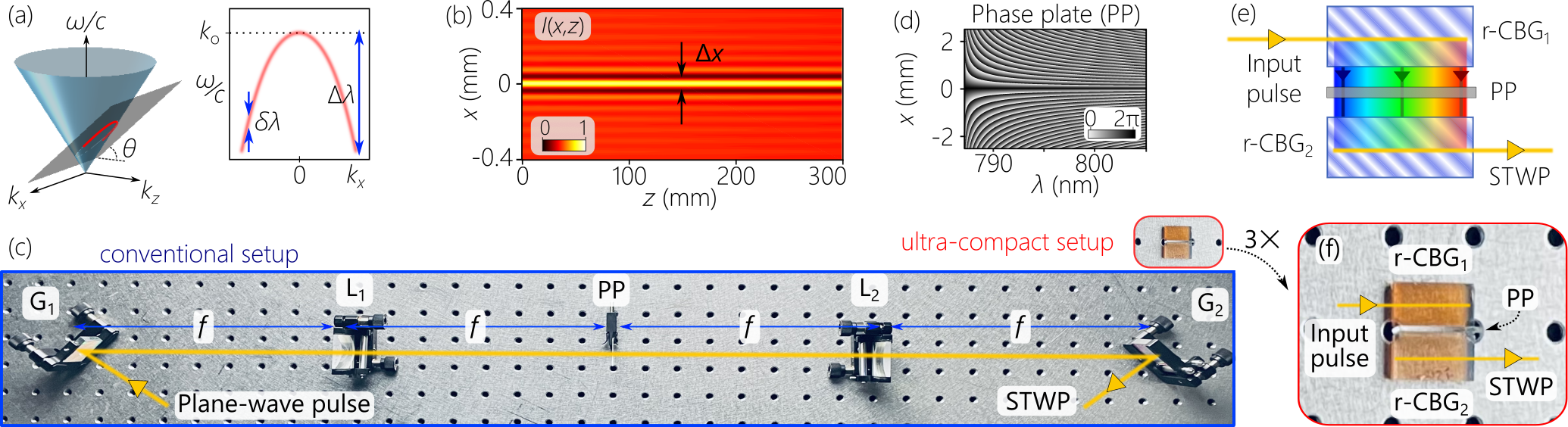}
    \caption{(a) Representation of the spatio-temporal spectrum of an STWP on the surface of the light-cone at its intersection with a tilted spectral plane. The spectral projection onto $(k_{x},\tfrac{\omega}{c})$-domain is a conic section. (b) Time-averaged axial intensity profile $I(x,z)$ of  an STWP with $\theta\!=\!44.97^{\circ}$ and beam size $\Delta x\!\approx\!40$~$\upmu$m. (c) Conventional free-space setup for producing STWPs and (d) the phase plate that introduces the requisiste spatio-temporal spectral structure into the field. (e) Schematic of the ultra-compact synthesis system for STWPs based on r-CBGs, and (f) its physical realization. Note that we also show the r-CBG-based system in (c) at the same scale as the conventional setup for comparison. G: Diffraction grating; L: cylindrical lens; PP: phase plate.}
    \label{fig:STconcept}
\end{figure*}

Such a device can help dramatically reduce the size of the conventional spatio-temporal synthesis setup for STWPs developed over the course of the past $\sim\!5$~years \cite{Yessenov22AOP}. The key feature of an STWP is that a particular form of angular dispersion -- a wavelength-dependent propagation angle $\varphi(\omega)$ -- is introduced into a generic plane-wave pulse, such that the axial wave number $k_{z}$ and temporal frequency $\omega$ are related via a linear relationship $\Omega\!=\!(k_{z}-k_{\mathrm{o}})c\tan{\theta}$ in free space, where $\Omega\!=\!\omega-\omega_{\mathrm{o}}$, $\omega_{\mathrm{o}}$ is a fixed frequency associated with on-axis propagation, $c$ is the speed of light in vacuum, $k_{\mathrm{o}}\!=\!\tfrac{\omega_{\mathrm{o}}}{c}$, and $\theta$ is the spectral tilt angle \cite{Kondakci17NP}. If we consider only one transverse coordinate $x$ (and hold the field uniform along $y$), then the spatio-temporal spectrum of the STWP is the conic section at the intersection of the light-cone $k_{x}^{2}+k_{z}^{2}\!=\!(\tfrac{\omega}{c})^{2}$ with a plane that is parallel to the $k_{x}$-axis and makes an angle $\theta$ with the $k_{z}$-axis [Fig.~\ref{fig:STconcept}(a)]; where $k_{x}$ is the transverse wave number along $x$. The STWP field $E(x,z;t)\!=\!e^{i(k_{\mathrm{o}}z-\omega_{\mathrm{o}}t)}\psi(x,z;t)$ has a propagation-invariant envelope $\psi(x,z;t)\!=\!\psi(x,0;t-z/\widetilde{v})$ that travels rigidly in free space at a group velocity $\widetilde{v}\!=\!c\tan{\theta}$ determined solely by the spectral tilt angle $\theta$ \cite{Kondakci19NC}, so the time-averaged intensity $I(x,z)\!=\!\int\!dt\,|E(x,z;t)|^{2}\!=\!I(x,0)$ is independent of $z$ [Fig.~\ref{fig:STconcept}(b)].

Synthesizing such an STWP requires an arrangement that (1) spatially resolves the spectrum of a generic pulse; (2) modulates the spatial distribution of each frequency $\omega$ to associate with it the requisite spatial frequency $k_{x}(\omega)$; before (3) recombining the spectrum to reconstitute the pulse. The first and final steps are typically realized with a diffraction grating, whereas the spatial modulation is achieved with a transmissive phase plate \cite{Kondakci18OE} or reflective spatial light modulator (SLM) in a folded system \cite{Kondakci17NP}. A photograph of a typical synthesis system that makes use of a transmissive phase plate is shown in Fig.~\ref{fig:STconcept}(c). The requisite phase distribution imparted by the transmissive phase plate to produce an STWP with $\theta\!=\!44.97^{\circ}$ over a bandwidth $\Delta\lambda\!\approx\!15$~nm is shown in Fig.~\ref{fig:STconcept}(d). The corresponding system using a reflective SLM would cut the volume in half.  

The spatio-temporal synthesis system based on r-CBGs makes use of the same strategy. However, rather than the combination of a diffraction grating and collimating lens (which occupy a large volume and require careful alignment), the resolved spectrum is formed at the output facet of r-CBG$_{1}$, the phase plate abuts that facet, and r-CBG$_{2}$ is placed immediately after the phase plate. In other words, the two r-CBGs sandwich the phase plate as shown in Fig.~\ref{fig:STconcept}(e). The two r-CBGs are designed to yield a linear spectral chirp of $\approx\!0.7$~nm/mm over a length of $L\!=\!25$~mm. To bring about clearly the dramatic reduction in footprint of this synthesis arrangement based on r-CBGs, we place it alongside the conventional setup in Fig.~\ref{fig:STconcept}(c); a magnified view is provided in Fig.~\ref{fig:STconcept}(f). The total volume of the new system is only $25\times25\times8$~mm$^{3}$. 

\begin{figure}[b!]
    \centering
    \includegraphics[width=8.6cm]{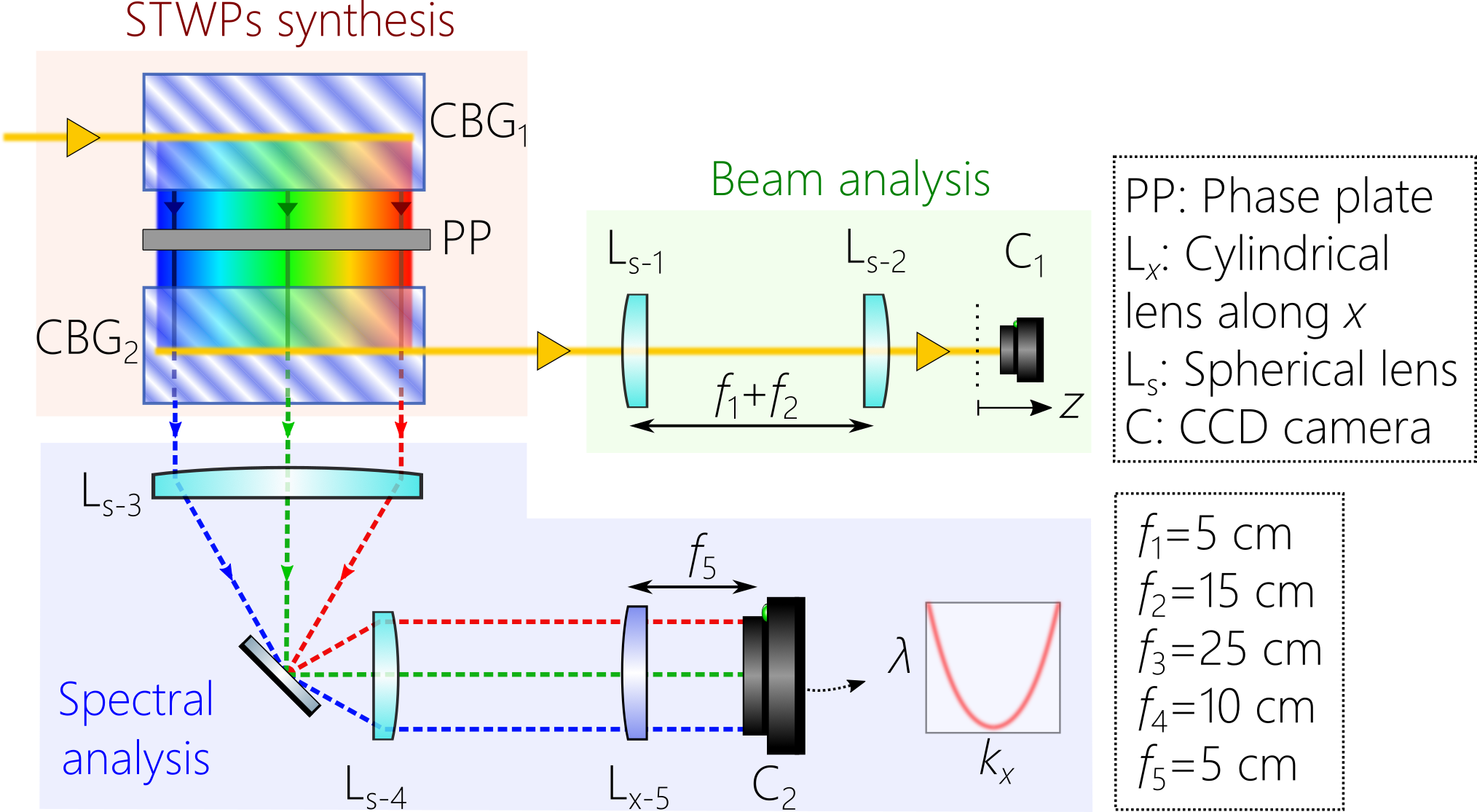}
    \caption{Layout of the setup for the ultra-compact synthesis of STWPs via a pair of r-CBGs and a phase plate, along with the characterization setup.}\vspace{-5mm}
    \label{fig:Setup}
\end{figure}

\begin{figure*}[t!]
    \centering
    \includegraphics[width=16cm]{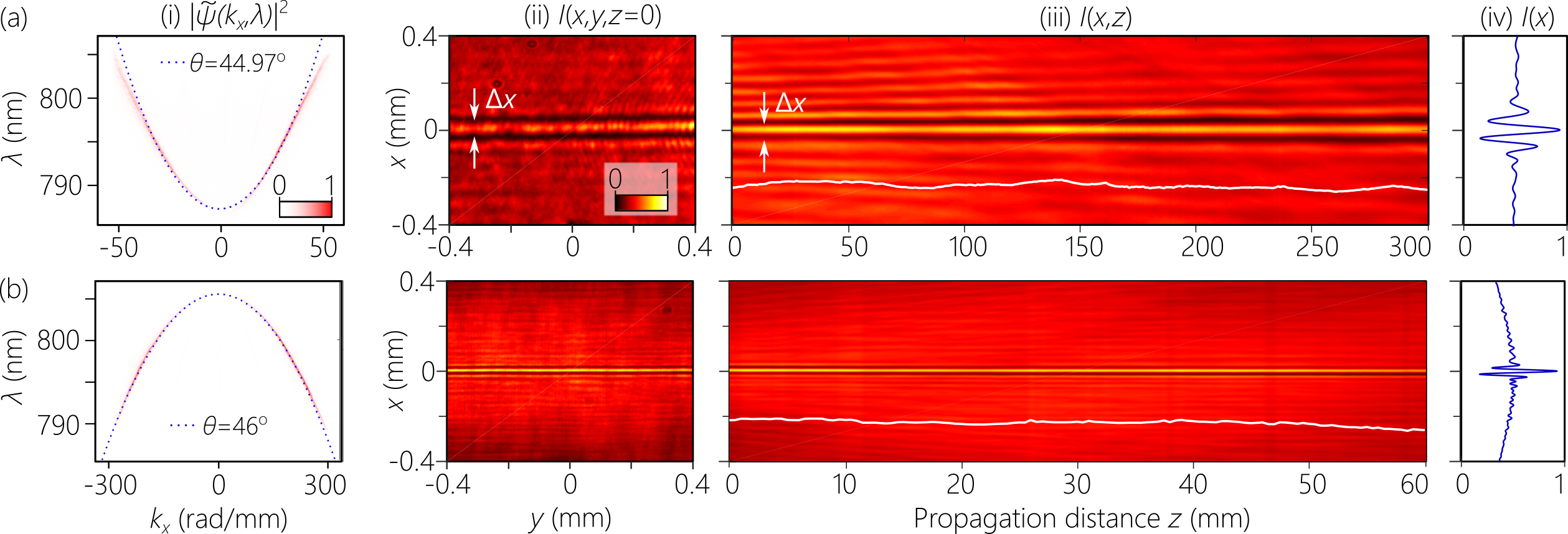}
    \caption{Diffraction-free propagation and spatio-temporal spectrum of the STWPs synthesized with the new system. (a) Subluminal STWP with $\theta\!=\!44.97^{\circ}$ and $\Delta x\!\approx\!40$~$\upmu$m; and (b) superluminal STWP with $\theta\!=\!46^{\circ}$ and $\Delta x\!\approx\!13$~$\upmu$m. The columns show the following quantities: (i) the spatio-temporal spectral intensity projected onto the $(k_{x},\lambda)$-plane; (ii) the time-averaged intensity in the transverse $(x,y)$-plane at $z\!=\!0$; (iii) the time-avaraged intensity $I(x,z)$, and (iv) $I(x,0)$.}
    \label{fig:TimeAveragedIntensity}
\end{figure*}

A detailed layout of the setup for the ultra-compact synthesis and the characterization of STWPs is depicted in Fig.~\ref{fig:Setup}. The pulses from Ti:Sapphire laser mentioned above are normally incident on the input facet of r-CBG$_{1}$, and the spatially resolved spectrum exits r-CBG$_{1}$ normally to the output facet along the $z$-axis. We place the transmissive phase plate [Fig.~\ref{fig:STconcept}(d)] abutting this facet to introduce the requisite spatio-temporal spectral structure associated with STWPs [Fig.~\ref{fig:STconcept}(a)]. The phase plate is fabricated using gray-scale lithography \cite{Mohammad17SR} and was designed to synthesize STWPs with tilt angle $\theta\!=\!46^{\circ}$ (superluminal STWP) over a bandwidth of $\Delta\lambda\!\approx\!17$~nm when oriented such that the longest wavelength is associated with $k_{x}\!=\!0$. When the phase plate is flipped horizontally so that $k_{x}\!=\!0$ is associated with the shortest wavelength in the spectrum, an STWP with $\theta\!=\!44^{\circ}$ (subluminal STWP) is produced. Placing r-CBG$_{2}$ immediately after the phase plate then reconstitutes the spectrum, and the beam exiting r-CBG$_{2}$ is an STWP with the desired spatio-temporal spectral structure. We place a telescope system in the path of the beam after r-CBG$_{2}$. For the superluminal STWP of spectral tilt angle $\theta\!=\!46^{\circ}$ in Fig.~\ref{fig:TimeAveragedIntensity}(b), the telescope system images the output plane of r-CBG$_{2}$ with unity magnification (the focal lengths of the two lenses are $f_{1}\!=\!f_{2}\!=5$~cm). For the subluminal STWP of spectral tilt angle $\theta\!=\!44^{\circ}$ in Fig.~\ref{fig:TimeAveragedIntensity}(a), the telescope introduces $3\times$ magnification ($f_{2}\!=\!3f_{1}\!=15$~cm), which changes the spectral tilt angle to $\theta\!=\!44.97^{\circ}$ and yields a larger beam width.

We characterize the synthesized STWP in the physical domain by scanning a CCD camera (C$_{1}$; TheImagingSource, DMK 27BUP031) along the propagation axis $z$ and monitor the axial evolution of the transverse spatial profile $I(x,z)$ [Fig.~\ref{fig:TimeAveragedIntensity}, columns (ii-iv)]. To confirm that the desired spatio-temporal spectral correlation between $k_{x}$ and $\lambda$ has been realized by the r-CBG-based setup, we characterize the STWPs in the spectral domain by measuring spatio-temporal spectrum $|\widetilde{\psi}(k_{x},\lambda)|^2$ projected onto the $(k_{x},\lambda)$-plane [Fig.~\ref{fig:TimeAveragedIntensity}, column (i)]. For that, we exploit the finite diffraction efficiency of r-CBG$_{2}$ and use the transmitted spatially resolved spectrum on which the phase plate has imprinted a spatio-temporal spectral structure. A CCD camera (C$_{2}$; TheImagingSource, DMK 33UX178) placed in the Fourier plane of a cylindrical lens $L_{x-5}$ captures the spatio-temporal spectrum. A telescope (focal lengths $f_{4}\!=\!f_{3}/2.5\!=10$~cm) placed in the path of the spectrum to de-magnifies the field by $2.5\times$ to fit the spatio-temporal spectrum into the active area of $C_{2}$.

The diffraction-free characteristics of the produced STWPs are confirmed in Fig.~\ref{fig:TimeAveragedIntensity}. We first plot the spatio-temporal spectra projected onto the $(k_{x},\lambda)$-plane for the subluminal [Fig.~\ref{fig:TimeAveragedIntensity}(a), column (i)] and superluminal [Fig.~\ref{fig:TimeAveragedIntensity}(a), column (i)] STWPs. Next we plot the time-averaged intensity profiles in the transverse plane at $z\!=\!0$ [Fig.~\ref{fig:TimeAveragedIntensity}, column (ii)] and along the optical axis $I(x,z)$ [Fig.~\ref{fig:TimeAveragedIntensity}, column (iii)]. The propagation length in Fig.~\ref{fig:TimeAveragedIntensity}(a) exceeds the Rayleigh range of a Gaussian beam with the same width $\Delta x$ by $50\times$, and in Fig.~\ref{fig:TimeAveragedIntensity}(b) by $90\times$.

We have drastically reduced the size of the setup needed for synthesizing STWPs by eliminating the need for diffraction gratings and collimating lenses. By utilizing instead r-CBGs, the STWP-synthesis system is compact ($\sim\!25\times25\times8$~mm$^{3}$), easily aligned, and robust. These newly developed r-CBGs can be similarly used to reduce the size of the synthesis arrangement for other spatio-temporally structured pulses. It remains an open question whether the system size can be further reduced by relying on a single metasurface or a pair of metasurfaces.

In conclusion, we have constructed an ultra-compact setup for synthesizing STWPs based on an alternative strategy for spectrally resolving the spectrum of an optical pulse. Rather than using conventional diffraction gratings, we employ chirped Bragg volume grating in which the Bragg structure is rotated by $45^{\circ}$ with respect to the plane-parallel facets of the device. Because the spectrum is resolved without further need of free-space propagation or collimation by a lens, and because the spatially resolved spectrum exits normally from a facet orthogonal to that of the input, we can sandwich a phase plate that spatially modulates the spectrally resolved wavefront between a cascade of two r-CBGs. The STWPs produced using this ultra-compact configuration have the same characteristics as those produced by setups based on diffraction gratings, whose volume exceeds that of the r-CBG-based setup by several orders of magnitude.

\textbf{Funding:}
U.S. Office of Naval Research (ONR) N00014-17-1-2458 and N00014-20-1-2789.\\

\textbf{Disclosures:}
The authors declare no conflicts of interest.\\

\bibliography{diffraction}


\end{document}